\documentclass[10pt,aps,prd,floatfix,nofootinbib,superscriptaddress,twocolumn]{revtex4-2}
\usepackage{graphicx}  
\usepackage{amsfonts,amsmath,amssymb,bm}        
\usepackage{multirow,bigdelim}
\usepackage[dvipsnames]{xcolor}
\usepackage[colorlinks,linkcolor=blue,citecolor=Green]{hyperref}
\usepackage[inline]{enumitem}
\usepackage[utf8]{inputenc}
\usepackage{orcidlink}

\newcommand{\itp}{\affiliation{Institute of Theoretical Physics, Chinese Academy of Sciences,\\ Zhong Guan Cun East Street 55, Beijing 100190, China}}

\newcommand{\ucas}{\affiliation{School of Physical Sciences, University of Chinese Academy of Sciences, Beijing 100049, China}}

\newcommand{\peng}{\affiliation{Peng Huanwu Collaborative Center for Research and Education, Beihang University, Beijing 100191, China}}

\newcommand{\scnt}{\affiliation{Southern Center for Nuclear-Science Theory (SCNT), Institute of Modern Physics,\\ Chinese Academy of Sciences, Huizhou 516000, China}}
\newcommand{\thu}{\affiliation{Department of Physics, Tsinghua University, Beijing 100084, China}}

\begin{document}

\title{
Evidence for the existence of a flavor-sextet charmed meson?
}

\author{Feng-Kun~Guo\orcidlink{0000-0002-2919-2064}}\email{fkguo@itp.ac.cn}
\itp \ucas \peng \scnt
\author{Bing-Song Zou\orcidlink{0000-0002-3000-7540}}\email{zoubs@mail.tsinghua.edu.cn}
\thu



\maketitle

The quest for understanding the internal structure of hadrons is key to probing how color confinement, which remains a fundamental challenge in the Standard Model, manifests itself.
The lowest-lying hadrons fit well into the classification of the conventional quark model, such as quark-antiquark mesons and three-quark baryons.
However, for excited states, the energy excitation inside a hadron can be distributed in various ways: it can appear as orbital or radial excitation within a conventional quark model configuration, or it leads to the creation of a quark-antiquark pair.
In the latter case, the multiquark configuration can still be arranged in different ways: either as compact multiquark states, where the confinement interaction occurs among all possible quark-(anti)quark pairs, or as hadronic molecules (with atomic nuclei as special examples), where the confinement interaction occurs only within the smallest possible color-singlet systems and the constituent hadrons interact via the residual strong force.
These different models are characterized by distinct SU(3) flavor structures. Therefore, identifying the complete multiplet family for a given set of spin-parity quantum numbers is crucial for determining the underlying structure.

Recently, the LHCb Collaboration reported a signal for a new resonance in the $D_s^+ \pi^\pm$ invariant mass distribution in the decays $B\to \bar{D}^{(*)} D_s^+ \pi^+ \pi^-$~\cite{LHCb:2024iuo}. This could be a direct observation of an SU(3) flavor-sextet charmed meson with $J^P=0^+$. Such an SU(3) multiplet, to be discussed below, is beyond the conventional quark-antiquark picture, and thus a verification of this observation is important. 

The year 2003 marked a renaissance in hadron spectroscopy. During that year, several influential observations were made: $\Theta(1540)$, $D_{s0}^*(2317)$, $D_{s1}(2460)$, and $X(3872)$. 
These discoveries sparked intensive theoretical investigations.
While the first candidate disappeared due to null signals in high-statistics experiments, the existence of the other states has been firmly established~\cite{ParticleDataGroup:2024cfk}.

The $D^{\ast}_{s0}(2317)$ and $D_{s1}(2460)$ possess spin-parity quantum numbers of $J^P=0^+$ and $J^P=1^+$, respectively. Their observed masses are significantly lower than those predicted by quark models for the lowest-lying charm-strange mesons with these quantum numbers.
Consequently, various models were proposed to explain these states, including $c\bar{s}$ states with unquenching effects (which incorporate mixing between the $c\bar{s}$ and tetraquark or meson-meson components), compact tetraquarks, and heavy-light meson-meson molecules.

The approximate SU(3) flavor symmetry of quantum chromodynamics (QCD) dictates that these states cannot exist in isolation. They must belong to flavor multiplets of the SU(3) group.
Regardless of the theoretical model employed, the charm-strange mesons should have non-strange SU(3) partners: the $D_0^*$ isospin doublet for $0^+$ mesons and the $D_1$ for $1^+$ mesons.
Since the strange quark is heavier than the up and down quarks, the $D_{s0}^*$ and $D_{s1}$ should be heavier than their non-strange counterparts $D_0^*$ and $D_1$ within the same SU(3) multiplet. 
This mass hierarchy is generally observed throughout the hadron spectrum; for example, the $K^*(892)$ is approximately 100 MeV heavier than the $\rho$ and $\omega$ mesons.
Surprisingly, the $D_0^*$ and $D_1$ resonances subsequently reported in experimental analyses have masses comparable to their charm-strange counterparts.
According to the Review of Particle Physics, the masses are $M_{D_0^*} = (2343\pm10)$~MeV and $M_{D_1} = (2412\pm9)$~MeV~\cite{ParticleDataGroup:2024cfk}, derived from averaging Breit-Wigner fits to $D\pi$ and $D^*\pi$ invariant mass distributions from various experiments.
These masses have been used as starting points in some phenomenological studies. However, it is important to note that the Breit-Wigner parameterization, which is effective for isolated narrow resonances far from strongly-coupled-channel thresholds, is inadequate for describing the $D_0^*(2300)$ and $D_1(2430)$ states. These resonances are broad, and their mass regions encompass the thresholds of $D\eta$, $D_s\bar{K}$ for $D_0^*$, and $D^*\eta$, $D_s^*\bar K$ for $D_1$, respectively. Furthermore, the amplitudes for reactions involving these resonances must satisfy chiral symmetry constraints~\cite{Du:2019oki}, a requirement that the Breit-Wigner parameterization fails to meet.

\begin{table}[t]
  \centering
  \caption{SU(3) flavor multiplets of positive-parity charmed mesons in different models.}
  \begin{tabular}{ll}
    \hline
    Models & SU(3) flavor multiplets  \\
    \hline
    $c\bar{q}$ (w/ or w/o unquenching effects) & $[\mathbf{\overline{3}}]$ \\
    Hadronic molecules & $[\mathbf{\overline{3}}]\oplus[\mathbf{6}]$ \\
    Diquark-antidiquark tetraquarks & $[\mathbf{\overline{3}}]\oplus[\mathbf{6}]\oplus[\mathbf{\overline{15}}]$ \\
    \hline
  \end{tabular}
  \label{tab:flavor_multiplets}
\end{table}

Let us consider the SU(3) flavor structure of different models for the positive-parity charmed mesons.
First, in the quark-antiquark picture, $c\bar q$ ($q=u,d,s$) states belong to the flavor anti-triplet. 
Introducing unquenching effects without further considering the interactions between the created $q\bar q$ pair or meson-meson components does not alter this group structure.
Second, in the hadronic molecular picture, the $D_{s0}^*(2317)$ and $D_{s1}(2460)$ are interpreted as $DK$ and $D^*K$ hadronic molecules, respectively~\cite{Barnes:2003dj, Chen:2004dy}. This naturally explains the near-equality between the mass differences $M_{D_{s1}(2460)} - M_{D^\ast_{s0}(2317)}$ and $M_{D^*} - M_{D}$ as a consequence of heavy quark spin symmetry.
The group theoretical structure of the hadronic molecular picture arises from the interaction between the flavor antitriplet $c\bar q$ mesons ($D$ and $D_s$) and the flavor octet light pseudoscalar mesons ($\pi$, $K$ and $\eta$ mesons): $[\mathbf{\overline{3}}]\otimes[\mathbf{8}]=[\mathbf{\overline{3}}]\oplus[\mathbf{6}]\oplus[\mathbf{\overline{15}}]$. However, since pions, kaons and $\eta$ are pseudo-Nambu-Goldstone bosons whose interactions with matter fields are constrained by chiral symmetry, the leading order chiral interaction is attractive only in the $[\mathbf{\overline{3}}]$ and $[\mathbf{6}]$ channels, with the former being the most attractive. As a result, only flavor anti-triplet and sextet states can form~\cite{Albaladejo:2016lbb}.  
Third, in the compact tetraquark picture based on the diquark-antidiquark model, in addition to the flavor anti-triplet and sextet~\cite{Maiani:2024quj}, the $[\mathbf{\overline{15}}]$ representation is also present when considering both spin-0 and spin-1 diquarks~\cite{Guo:2025imr}. The flavor multiplets of these various models are summarized in Table~\ref{tab:flavor_multiplets}.
Therefore, experimental determination of the flavor structure of positive-parity charmed mesons is crucial for distinguishing between these theoretical models.

\begin{figure}[tb]
  \centering
  \includegraphics[width= \linewidth]{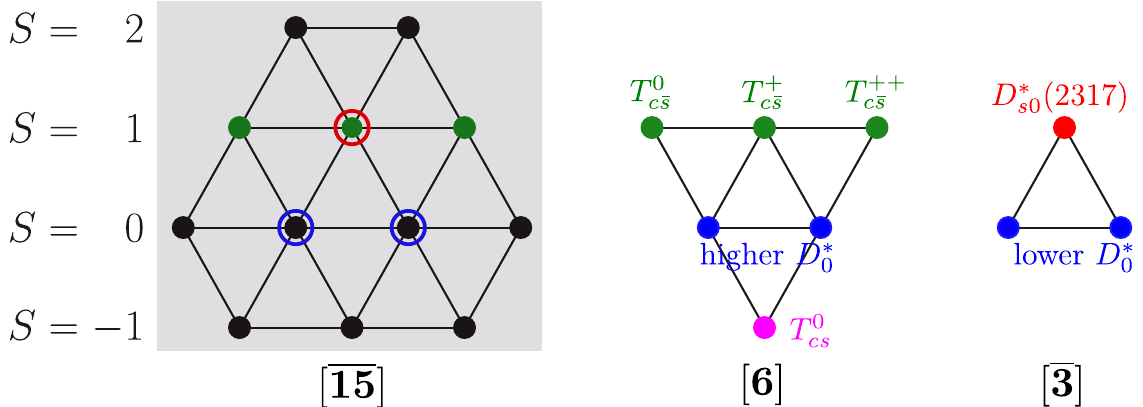} 
  \caption{Decomposition of the charmed-meson--light-meson pairs into different SU(3) flavor multiplets. }
  \label{fig:irreps}
\end{figure}
Quantitative predictions in the hadronic molecular picture can be made using unitarized chiral perturbation theory (UChPT), with inputs from lattice QCD, regarding the masses, decays, and interactions between charmed mesons and light mesons.
Since the leading order $S$-wave chiral interaction, given by the Weinberg-Tomozawa term, is proportional to the energy of the light meson, the interaction strength at low energies is sensitive to the masses of the involved light mesons. 
As kaons are considerably heavier than pions, the $S$-wave $DK$ interaction is substantially stronger than the $D\pi$ interaction. Consequently, the $D_{s0}^*$ ($D_{s1}$) appears as a bound state in the $DK$ ($D^*K$) channel (a scenario supported by analyses of lattice QCD data (for a review, see Ref.~\cite{Guo:2023wkv}),
while the $D\pi$ ($D^*\pi$) channel produces only a broad resonance above the threshold. In fact, $I=1/2$ isospin doublets exist in all three multiplets, as shown by the blue dots and circles in Fig.~\ref{fig:irreps}, and their mixture leads to two $D_0^*$ resonances~\cite{Kolomeitsev:2003ac,Guo:2006fu,Albaladejo:2016lbb}.
The lower $D_0^*$ resonance is the SU(3) partner of the $D_{s0}^*(2317)$ in the anti-triplet, while the higher $D_0^*$ resonance originates from the sextet.
It has been demonstrated in Refs.~\cite{Du:2017zvv,Du:2019oki} that the precise measurements by the LHCb Collaboration of the angular moments of the $B^-\to D^+ \pi^- \pi^-$~\cite{LHCb:2016lxy} and related three-body decays can be well described by the UChPT amplitudes that include both $D_0^*$ resonances.
Lattice support for the higher $D_0^*$ pole in the flavor sextet at a large SU(3) symmetric pion mass, as suggested in Ref.~\cite{Du:2017zvv}, has recently been reported in Ref.~\cite{Yeo:2024chk}.
It was also predicted in Ref.~\cite{Albaladejo:2016lbb} that the sextet contains an isoscalar $D\bar K$ virtual state ($T_{cs}^0$ in Fig.~\ref{fig:irreps}), and this prediction was later confirmed by lattice calculations in Ref.~\cite{Cheung:2020mql}. Such a virtual state is a pure sextet state as it does not receive any mixture from other multiplets.
The situation for the isospin triplet in the flavor sextet, however, is more complex since it mixes with the isovector in the $[\mathbf{\overline{15}}]$ multiplet that has a repulsive interaction.
Accordingly, the isospin-1 $DK$-$D_s\pi$ coupled-channel scattering amplitude is expected to have a pole deep in the complex plane. It was predicted to have a real part of about 2.3~GeV and a large imaginary part between 0.1 and 0.2~GeV on the third Riemann sheet~\cite{Guo:2009ct} (the prediction needs to be updated with the parameters determined later in Ref.~\cite{Liu:2012zya}). Such a pole would manifest in the $D_s\pi$ invariant mass distribution only as a moderate cusp at the $DK$ threshold, making it difficult to observe experimentally.

Despite substantial support from lattice QCD calculations for the hadronic molecular picture, direct experimental evidence has remained elusive. 
Beyond the flavor structure emphasized above, 
a compelling test of the hadronic molecular nature of the $D_{s1}(2460)$ involves measuring the dipion invariant mass distribution in the $D_{s1}(2460)\to D_s \pi^{+} \pi^{-}$ decay.
It was predicted in Ref.~\cite{Tang:2023yls} that this distribution would exhibit a distinctive double-bump structure if the $D_{s1}(2460)$ possesses a hadronic molecular structure, in contrast to a single broad bump expected for a compact state.
This characteristic double-bump line shape arises from two mechanisms: contributions from loops involving the $D^*K$ intermediate state (to which the $D_{s1}(2460)$ couples strongly as a defining feature of the molecular model) that enhance the lower end of the spectrum, and the $\pi\pi$ final state interaction (primarily through the $f_0(500)$ scalar meson) that enhances the higher end of the distribution.
Experimental confirmation of this double-bump structure would provide strong evidence for the hadronic molecular nature of the $D_{s1}(2460)$, and by extension, support the molecular picture for the $D_{s0}^*(2317)$ as a natural consequence of heavy quark spin symmetry.

Such a measurement has been performed by LHCb as reported in Ref.~\cite{LHCb:2024iuo}, where a double-bump structure in the $\pi^+\pi^-$ invariant mass distribution of the $D_{s1}(2460)\to D_s \pi^+\pi^-$ decay was indeed observed.
The Dalitz plot distribution of the decay was analyzed using two different models: one incorporating $f_0(500)$, $f_0(980)$, and $f_2(1270)$ resonances, and another considering $f_0(500)$ along with additional scalar $T_{c\bar s}$ resonances coupled to $D_s\pi^\pm$. 
Although both models provided a good description of the data, the former was disfavored because the extracted mass and width of the $f_0(500)$ were significantly smaller than the established values~\cite{ParticleDataGroup:2024cfk}.
In the latter model, the $T_{c\bar s}$ was determined to have a mass of $(2327 \pm 13 \pm 13)$~MeV and a width of $\left(96 \pm 16_{-\phantom{1}23}^{+170}\right)$~MeV, with scalar $J^P=0^+$ quantum numbers being favored over vector ones.
It is worth noting that even in this latter model, the width of the $f_0(500)$ from the fit, 0.2~GeV, remains considerably smaller than its well-established value.

The $T_{c\bar s}$ resonance, if confirmed, would be a natural candidate for the isovector in the flavor sextet, as illustrated in Fig.~\ref{fig:irreps}.
In addition to the LHCb analysis, the experimental data for the $\pi\pi$ and $D_s\pi$ invariant mass distributions have been well reproduced through model fits in Refs.~\cite{Wang:2024fsz,Roca:2025lij}. 
Notably, while Ref.~\cite{Wang:2024fsz} incorporated a scalar $T_{c\bar s}$ resonance through dynamical generation, Ref.~\cite{Roca:2025lij} treated the $DK\to D_s\pi$ transition perturbatively (thus no $T_{c\bar s}$ resonance was included), similar to the approach in Ref.~\cite{Tang:2023yls}.
However, since the UChPT amplitude for the isovector $DK$-$D_s\pi$ coupled-channel scattering does contain a pole---albeit deep in the complex energy plane---as predicted in Ref.~\cite{Guo:2009ct}, a nonperturbative treatment would be more appropriate for describing the $DK\to D_s\pi$ scattering.
Furthermore, the $\pi\pi$ final state interaction should be quantitatively consistent with precisely determined scattering data, as implemented in Ref.~\cite{Tang:2023yls}, which is not achieved in Ref.~\cite{Wang:2024fsz}.
Therefore, considering the significance of this experimental observation in resolving the longstanding puzzle of positive-parity charmed mesons, a comprehensive analysis is essential to verify the existence of the isovector $T_{c\bar s}$ resonance.

Furthermore, the scalar $T_{c\bar s}$ resonance should have a heavy quark spin partner with $J^P=1^+$. Such an $T_{c\bar s1}$ resonance should show up as a moderate cusp at the $D^*K$ threshold in the hadronic molecular picture, and it can be searched for in $D_s^*\pi^\pm$ invariant mass distributions of, e.g., the $B\to \bar D^{(*)} D_s^*\pi^\pm(\pi)$ decays.

\begin{acknowledgments}
  This work is supported in part by the National Natural Science Foundation of China under Grants No. 12125507, No. 12361141819, and No. 12047503; by the National Key R\&D Program of China under Grant No. 2023YFA1606703; and by the Chinese Academy of Sciences under Grant No.~YSBR-101. 
  \end{acknowledgments}

\bibliography{refs}

\end{document}